\documentclass[3p, twocolumn, number]{elsarticle}

\usepackage[utf8]{inputenc}
\usepackage{textcomp}
\usepackage{multirow}
\usepackage{nidanfloat}
\usepackage{graphicx}
\usepackage{mathtools}
\usepackage{amssymb}
\usepackage{amsmath}
\usepackage{bm}
\usepackage{subcaption}
\usepackage{siunitx}
\usepackage{longtable,tabularx}
\usepackage{float}
\usepackage{booktabs}
\usepackage{bookmark}
\usepackage{etoolbox}
\usepackage[title]{appendix}
\usepackage{subcaption}
\usepackage[dvipsnames]{xcolor}

\usepackage{url}
\usepackage{hyperref}
\usepackage{bernsteinStyle}

\usepackage{url}
\usepackage{tikz}
\usepackage{pgfplots}
\usepackage{accents}
\usepackage{comment}

\newcommand{\abs}[1]{\left\vert #1\right\vert }
\newcommand\atopJ[2]{\genfrac{}{}{0pt}{}{#1}{#2}}
\usetikzlibrary{calc}
\usetikzlibrary{shapes, arrows}
\usetikzlibrary{positioning}
\usepackage{color}
\usepackage{circuitikz}
\usepackage{etoolbox}
\makeatletter
\patchcmd\@combinedblfloats{\box\@outputbox}{\unvbox\@outputbox}{}{\errmessage{\noexpand patch failed}}
\makeatother

\usetikzlibrary{shapes,arrows,calc,positioning}
\tikzstyle{bigblock} = [draw, fill=blue!20, rectangle, 
    minimum height=6em, minimum width=8em]
\tikzstyle{medblock} = [draw, fill=blue!20, rectangle, 
    minimum height=4em, minimum width=4em]    
\tikzstyle{mux} = [draw, fill=black!20, rectangle, 
    minimum height=5em, minimum width=0.1em]    
\tikzstyle{smallblock} = [draw, fill=blue!20, rectangle, 
    minimum height=2.5em, minimum width=4em]
\tikzstyle{sum} = [draw, fill=blue!20, circle, node distance=1cm]
\tikzstyle{signal} = [coordinate]
\tikzstyle{pinstyle} = [pin edge={to-,thin,black}]
\tikzstyle{block} = [draw, fill=blue!20, rectangle, 
    minimum height=3em, minimum width=6em]
\tikzstyle{blockS} = [draw, fill=blue!20, rectangle, 
    minimum height=3em, minimum width=4em]  
\tikzstyle{sum} = [draw, fill=blue!20, circle, node distance=1.5cm]
\tikzstyle{gain} = [draw, fill=blue!20, regular polygon, regular polygon sides = 3, node distance=1.25cm, shape border rotate = -90]
\tikzstyle{mult} = [draw, fill=blue!20, circle, inner sep=0pt, minimum size=0.2cm,]
\tikzstyle{saturation block} = [draw, fill=blue!20,
    	path picture={
    		\pgfpointdiff{\pgfpointanchor{path picture bounding box}{north east}}%
    		{\pgfpointanchor{path picture bounding box}{south west}}
    		\pgfgetlastxy\x\y
    		\tikzset{x=\x*.4, y=\y*.4}
    		\draw (-1,0) -- (1,0) (0,-1) -- (0,1); 
    		\draw (-1,-.7) -- (-.5,-.7) -- (.5,.7) -- (1,.7);
    	}]
\tikzstyle{sat atan} = [draw, fill=blue!20, 
        path picture={
          \pgfpointdiff{\pgfpointanchor{path picture bounding box}{north east}}%
            {\pgfpointanchor{path picture bounding box}{south west}}
          \pgfgetlastxy\x\y
          \tikzset{x=\x*.05, y=\y*0.3}
          \draw (-7.5,0) -- (7.5,0) (0,-1.2) -- (0,1.2); 
          \draw[scale = 1,domain=-7.5:7.5,smooth,variable=\x,black] plot ({\x},{(tanh((0.6*\x)))});
        }]
\tikzstyle{input} = [coordinate]
\tikzstyle{output} = [coordinate]

\pgfplotsset{compat=1.15}

\begin{document}

\begin{frontmatter}

\title{Self-Excited Dynamics of Discrete-Time Lur'e Systems}

\author[1]{Juan A. Paredes\corref{cor1}}
\ead{jparedes@umich.edu}

\author[1]{Syed Aseem Ul Islam}
\ead{aseemisl@umich.edu}

\author[2]{Omran Kouba}
\ead{omran_kouba@hiast.edu.sy}

\author[1]{Dennis S. Bernstein}
\ead{dsbaero@umich.edu}

\cortext[cor1]{Corresponding author}

\address[1]{University of Michigan, Ann Arbor, Michigan, 48109, USA}
\address[2]{Higher Institute for Applied Sciences and Technology, Damascus, Syria}

\begin{abstract}
Self-excited systems arise in numerous applications, such as biochemical systems, fluid-structure interaction, and combustion. 
This paper analyzes a discrete-time Lur'e system with a piecewise-linear saturation feedback nonlinearity.
The main result provides sufficient conditions under which the Lur'e system is self-excited in the sense that its response is bounded and nonconvergent.   
\end{abstract}

\end{frontmatter}

\section{Introduction}

A self-excited system has the property that the input is constant but the response is oscillatory.
Self-excited systems arise in numerous applications, such as biochemical systems, fluid-structure interaction, and combustion.  
The classical example of a self-excited system is the van der Pol oscillator, which has two states whose asymptotic response converges to a limit cycle.
A self-excited system, however, may have an arbitrary number of states and need not possess a limit cycle.
Overviews of self-excited systems are given in \cite{Ding2010,JENKINS2013167}.

Models of self-excited systems are typically based on the relevant physics of the application. 
From a systems perspective, the main interest is in understanding the features of the components of the system that give rise to self-sustained oscillations.
Understanding these mechanisms can illuminate the relevant physics in specific domains and provide unity across different domains.

A unifying model for self-excited systems is a feedback loop involving linear and nonlinear elements; systems of this type are called {\it Lur'e systems}. 
Lur'e systems have been widely studied in the classical literature on stability theory \cite{khalil3rd}.
Within the context of self-excited systems, Lur'e systems under various assumptions are considered in \cite{chua1979,Tomberg1989,hang2002,jian2004,gstan2007,aguilar2009,Ding2010,CHATTERJEE20111860,Gusman_2016,Zanette_2017}.
Self-oscillating discrete-time systems are considered in \cite{vrasvan98,amico2011}.

Roughly speaking, self-excited oscillations arise from a combination of stabilizing and destabilizing effects.
Destabilization at small signal levels causes the response to grow from the vicinity of an equilibrium, whereas stabilization at large signal levels causes the response to decay when the state is far from an equilibrium.
In particular, negative damping at low signal levels and positive damping at high signal levels is the mechanism that gives rise to a limit cycle in the van der Pol oscillator \cite[pp. 103--107]{nayfeh2008}.
Note that, although systems with limit-cycle oscillations are self-excited, the converse need not be true since the response of a self-excited system may oscillate without the trajectory reaching a limit cycle.
Alternative mechanisms exist, however; for example, time delays are destabilizing, and Lur'e models with time delay have been considered as models of self-excited systems \cite{minorskypaper}.

The present paper considers a discrete-time Lur'e system with asymptotically stable linear dynamics, a  zero at 1, and a piecewise-linear saturation feedback nonlinearity.
For this Lur'e system, sufficiently large scalings of the asymptotically stable dynamics yield closed-loop unstable dynamics while the saturation function operates in its linear region.
Under large signal levels, the saturation function yields a constant signal, which effectively breaks the loop, thus allowing the open-loop dynamics to stabilize the response.
The  zero at 1 acts as a high-pass filter, which ensures that the response does not converge, whereas the saturation function yields a constant signal.
Hence, while the saturation function operates in the nonsaturated region, the closed-loop system is unstable, and, while the saturation function operates in the saturated region and yields a constant signal, the closed-loop system is asymptotically stable and has a constant input.
The contribution of the present paper is to prove that this model structure  yields self-excited oscillations for sufficiently large scalings of the asymptotically stable dynamics.
A preliminary study of self-excited oscillations in a similar discrete-time Lur'e model was performed in \cite{DTLACC}.
However, the present paper goes far beyond \cite{DTLACC} in breadth and depth of the analysis of these systems.

The contents of the paper are as follows. 
Section \ref{sec:TFfeedback} considers a simple discrete-time linear feedback system and analyzes its transfer function to study the range of values of the linear dynamics scalings for which the closed-loop system is asymptotically stable.
Section \ref{sec:SSfeedback} considers the same linear feedback system and analyzes its state space model to study the conditions under which the response of the closed-loop system does not converge and is not bounded.
Section \ref{sec:lure} extends the problem in Sections \ref{sec:TFfeedback} and \ref{sec:SSfeedback} by including a saturation nonlinearity. Under certain conditions, this discrete-time Lur’e system is shown to have a bounded, non-convergent response for sufficiently large values of the loop gain.
Section \ref{sec:numericalEx} presents an example that illustrates the conditions for self-excitation presented in Section \ref{sec:lure}.

{\bf Nomenclature.}  $\BBR \isdef (-\infty, \infty),$ $\BBC$ denotes the complex numbers,
$\SR$ denotes range, $\SN$ denotes null space, $\overline{(\cdot)}$ denotes complex conjugate, $(\cdot)^*$ denotes complex conjugate transpose, $\Vert\cdot\Vert$ denotes the Euclidean norm on $\BBC^n,$ and $\bfq$ denotes the forward shift operator.
For a polynomial $p,$ define $\spr(p)\isdef\max\{|z|\colon z\in\BBC \mbox{ and } p(z) = 0\}$, and for a matrix $A\in\BBR^{n\times n},$ let $\spec(A)$ denote the set of eigenvalues of $A,$ let $\chi_A$ denote the characteristic polynomial of $A,$ and let $\spr(A)$ denote the spectral radius of $A.$
For $\gamma > 0,$  ${\rm sat}_\gamma$ denotes the saturation function, such that, for all $x \in [-\gamma, \gamma],$ ${\rm sat}_\gamma (x) = x,$ and, for all $|x|  >  \gamma,$  ${\rm sat}_\gamma (x) = (\sign x) \gamma.$

\section{Transfer Function Analysis of the Linear Feedback System} \label{sec:TFfeedback}

Let $G$ be a  strictly proper, asymptotically stable, discrete-time SISO transfer function with a  zero at 1 and no other zeros on the unit circle.
Let $G=N/D$, where the polynomials $N$ and $D$ are coprime, $D$ is monic, $n\isdef \deg D$, and $m\isdef \deg N < n.$
Note that $N(1) = 0$ and $D(1)\ne0,$ and thus $G(1) = 0.$
Furthermore, for all $\theta \in (-\pi,\pi]\backslash\{0\},$ $G(e^{\jmath\theta})\ne0.$

\begin{figure}[ht!]
    \centering
     \resizebox{0.5\columnwidth}{!}{%
    \begin{tikzpicture}[>={stealth'}, line width = 0.25mm]
    \node [input, name=input] {};
    \node [sum, right = 0.5cm of input] (sum1) {};
    \node[draw = white] at (sum1.center) {$+$};
    \node [smallblock,  rounded corners, right = 0.4cm of sum1, minimum height = 0.6cm, minimum width = 0.8cm] (system) {$\alpha  G(\bfq)$};
    \draw [->] (input) -- node[name=usys, xshift = -0.2cm, yshift = 0.2cm] {\scriptsize$v$} (sum1.west);
    \draw[->] (sum1.east) -- node [near end, above, xshift = -0.12cm] {\scriptsize$\nu$} (system.west);
    \node [output, right = 0.5cm of system] (output) {};
    \draw [->] (system) -- node [name=y,near end, xshift = -0.15cm]{} node [near end, above, xshift = 0.1cm] {\scriptsize$y$}(output);
    \node [input, below = 0.25cm of system] (midpoint) {};
    \draw [->] (y.center) |- (midpoint) -| (sum1.south);
    \end{tikzpicture}
    }
    \caption{\footnotesize Discrete-time linear feedback system with input $v$ and output $y$.}
    \label{DT_blk_diag}
\end{figure}
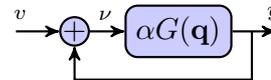

For all $\alpha\in\BBR,$ the closed-loop transfer function from $v$ to $y$ of the linear feedback system in Fig. \ref{DT_blk_diag} is given by
\begin{equation} \label{GyvEq}
   G_{yv,\alpha}(\bfq) \isdef \frac{\alpha G(\bfq)}{1 - \alpha G(\bfq)} = \frac{\alpha N(\bfq)}{p_{\alpha}(\bfq)},
\end{equation}
where $p_{\alpha}(\bfq) \isdef D(\bfq) - \alpha N(\bfq)$.
The forward shift operator $\bfq$ accounts for both the free and forced response of the linear feedback system in Fig. \ref{DT_blk_diag}; for pole-zero analysis, $\bfq$ is replaced by the Z-transform complex variable $\bfz$.
Note that $p_0 = D,$ and thus $\spr(p_0)<1.$
However, for all $|\alpha|$ sufficiently large, it follows from the root locus asymptote rule that $p_\alpha$ has at least $n-m$ roots outside the closed unit disk, and thus $\spr(p_\alpha)>1.$
Note that, since Fig. \ref{DT_blk_diag} has no sign change in the loop, the root locus parameter $\alpha$ plays the role of $-k$ in the standard root locus.
The following result is immediate.

\indent\begin{prop} \label{factAlpha}
Let $\alpha\in\BBR\backslash\{0\}$ and $\theta \in (-\pi, \pi] \backslash \{0\}$.
Then $p_{\alpha}(e^{\jmath \theta}) = 0$ if and only if  $\alpha = 1/G(e^{\jmath \theta}).$
\end{prop}

Proposition  \ref{factAlpha} implies that, if $\theta\in(-\pi, \pi]  \backslash \{0\}$ and $G(e^{\jmath \theta})$ is real, then $e^{\jmath\theta}$ is a pole of $G_{yv,1/G(e^{\jmath \theta})}$ and thus an element of either the 0-deg or 180-deg root locus of $G_{yv,\alpha}$.
Now, define  
\begin{align}
    \Theta_\rmn  &\isdef \{\theta \in(-\pi, \pi]  \backslash \{0\} \colon  G(e^{\jmath \theta}) \in (-\infty,0)  \},\\
    \Theta_\rmp  &\isdef \{\theta \in(-\pi, \pi] \backslash \{0\} \colon  G(e^{\jmath \theta}) \in(0,\infty)  \},
\end{align}
so that
\small
$\Theta_\rmn \ \cup \ \Theta_\rmp = \{\theta\in(-\pi, \pi] \backslash \{0\}\colon     G(e^{\jmath\theta}) \mbox{ is real}\}.$
\normalsize
Note that $\Theta_\rmn$ is the set of angles at which the 180-deg root locus of $G_{yv,\alpha}$ crosses the unit circle, and $\Theta_\rmp$ is the set of angles at which the 0-deg root locus of $G_{yv,\alpha}$ crosses the unit circle, which occurs in both cases for $\alpha = 1/G(e^{\jmath\theta}).$
Since the 180-deg and 0-deg root locus plots of $G_{yv,\alpha}$ have $n-m$ asymptotes as $\alpha\to-\infty$ and $\alpha\to\infty$, respectively, it follows that 
\begin{align}
    \card(\Theta_\rmn) &\ge n-m,\\ 
    \card(\Theta_\rmp) &\ge n-m.
\end{align}
Furthermore, in the case where $n-m=1,$ the positive real axis is an asymptote of the 0-deg root locus plot.
Since $G$ has a zero at 1, it follows that two poles must break in on the positive real axis, which implies that
\begin{equation}
    \card(\Theta_\rmp) \ge \min\{2,n-m\},
\end{equation}
as illustrated by the following example.

\begin{figure}[h]
    \centering
    \includegraphics[width=0.7\columnwidth]{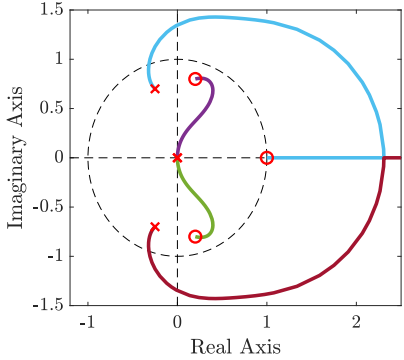}
    \caption{Example \ref{exam:alpha}: 0-deg root locus of the linear feedback system in Fig. \ref{DT_blk_diag}.}
    \label{fig:Ex_alpha}
\end{figure}

\begin{exam}\label{exam:alpha}
Let $\alpha \ge 0$ and $G(\bfq) = \tfrac{(\bfq - 0.2 \pm \jmath 0.8 )(\bfq - 1)}{\bfq^2 (\bfq + 0.25 \pm \jmath 0.7)} 
= \tfrac{(\bfq^2 - 0.4 \bfq + 0.68 )(\bfq - 1)}{\bfq^2 (\bfq^2 + 0.5 \bfq + 0.5525 )}.$ 
Since $n=4$ and $m=3,$ it follows that the 0-deg root locus of the closed-loop system has one asymptote, as shown in Fig. \ref{fig:Ex_alpha}. 
However, the root locus plot crosses the unit circle at two points due the pole break-in on the positive real axis, and thus ${\rm card}(\Theta_\rmp) = 2.$
\end{exam}

The following result implies that the root locus plot of $G$ intersects the unit circle at a finite number of points.

\indent\begin{prop}\label{propTheta}
$\Theta_\rmp$ and $\Theta_\rmn$ are finite. 
\end{prop}

{\bf Proof.}  
Let $r$ be a positive integer such that
	$h(\bfz)\isdef \bfz^r[D(\bfz)N(1/\bfz)-N(\bfz)D(1/\bfz)]$
is a polynomial.  
Now, let $\bfz= e^{\jmath\theta}$, where $\theta\in\Theta_\rmn\cup \Theta_\rmp.$  Since $G(\bfz)$ is real and $|\bfz|=1$, it follows that
\begin{equation*} 
    \frac{D(\bfz)}{N(\bfz)} = \overline {   \left( \frac{D(\bfz)}{N(\bfz)} \right)  }
    = {\frac{\overline D(\bfz)}{\overline N(\bfz)}} 
    =\frac{D(\bar{\bfz})}{N(\bar{\bfz})}=\frac{D(1/\bfz)}{N(1/\bfz)}.
\end{equation*}
Hence, $h(\bfz)=0.$  
Since $h$ has a finite number of roots, it follows that $\Theta_\rmn$ and $\Theta_\rmp$ are finite. 
\hfill $\square$

 Proposition \ref{propTheta} implies that we can define 
\begin{align}
    \alpha_\rmn &\isdef \max_{\theta \in \Theta_{\rmn}} 1/G(e^{\jmath \theta})<0, \label{alphanDef}\\
    \alpha_\rmp &\isdef \min_{\theta \in \Theta_{\rmp}}   1/G(e^{\jmath \theta})>0.\label{alphapDef}
\end{align}

\indent\begin{prop}\label{propGyvalphaAS}
If  $\alpha \in (\alpha_\rmn, \alpha_\rmp),$ then $G_{yv,\alpha}$ is asymptotically stable.
Furthermore,
\begin{align} \label{eqSpr1}
    \spr(p_{\alpha_\rmn}) = \spr(p_{\alpha_\rmp}) = 1.
\end{align}
\end{prop}

{\bf Proof.}
Suppose there exists $\alpha \in (\alpha_\rmn, 0)$ such that $\spr (p_\alpha) \geq 1.$
Since $\spr$ is continuous and $\spr (p_0) = \spr (D) < 1,$ the intermediate value theorem implies that there exists $\alpha_1 \in (\alpha_\rmn,0)$ such that  $\spr (p_{\alpha_1}) = 1.$
Hence, there exists $\theta_1 \in (-\pi,\pi]/\{0\}$ such that   $p_{\alpha_1}(e^{\jmath \theta_1})=0.$
Hence, $G(e^{\jmath \theta_1}) = 1/\alpha_1 < 0,$ and thus $\theta_1 \in \Theta_\rmn.$ 
Therefore, \eqref{alphanDef} implies that
\begin{equation*}
     \max_{\theta \in \Theta_{\rmn}} 1/G(e^{\jmath \theta}) = \alpha_\rmn  < \alpha_1 = 1/G(e^{\jmath \theta_1}),
\end{equation*}
which is a contradiction.  
Hence, for all $\alpha \in (\alpha_\rmn, 0],$ $G_{yv, \alpha}$ is asymptotically stable.
Similarly, for all $\alpha \in [0, \alpha_\rmp),$ $G_{yv, \alpha}$ is asymptotically stable.
Hence, for all $\alpha \in (\alpha_\rmn, \alpha_\rmp),$ $G_{yv, \alpha}$ is asymptotically stable.

Next, let $\theta_\rmn \in \Theta_\rmn$ satisfy $\alpha_\rmn = 1/G(e^{\jmath \theta_\rmn}).$
Then, Proposition \ref{factAlpha} implies that $p_{\alpha_\rmn} (e^{\jmath \theta_\rmn}) = 0,$ that is, $e^{\jmath \theta_\rmn}$ is a root of $p_{\alpha_\rmn},$ and thus $\spr (p_{\alpha_\rmn}) \geq 1.$ 
Now, suppose that $\spr (p_{\alpha_\rmn}) > 1.$
Since $\spr (p_0) < 1$ and $\spr$ is continuous, it follows that there exists $\alpha_1 \in (\alpha_\rmn,0)$ such that  $\spr (p_{\alpha_1}) = 1.$ 
Since $\alpha_1 \in (\alpha_\rmn, \alpha_\rmp),$ it follows that $G_{yv,\alpha_1}$ is asymptotically stable, and thus $\spr (p_{\alpha_1}) < 1,$ which is a contradiction. 
Hence, $\spr (p_{\alpha_\rmn}) = 1.$
Similarly, $\spr (p_{\alpha_\rmp}) = 1.$
\hfill $\square$

The following result is an immediate consequence of the root locus asymptote rule.

\begin{lem}\label{RLasy}
There exist $\beta_\rmn<0$ and $\beta_\rmp > 0$ such that, for all $\alpha<\beta_\rmn$ and all $\alpha>\beta_\rmp,$
$p_\alpha$ has at least $n-m$ roots with absolute value greater than 1, all of which are simple.
\end{lem}

The following example shows that $\alpha_\rmp$ defined by \eqref{alphapDef} is not necessarily the supremum of all values of $\alpha$ such that $G_{yv,\alpha}$ is asymptotically stable.
In other words, there may exist $\alpha>\alpha_\rmp$ such that $G_{yv,\alpha}$ is asymptotically stable.

\begin{figure}[ht!]
\centering
    \includegraphics[width=\columnwidth]{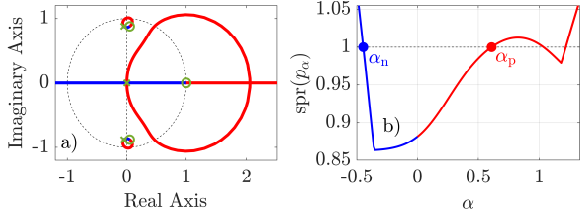}
    \caption{Example \ref{exam:spr_alpha}: Root locus and $\spr(p_\alpha)$ for the linear feedback system in Fig. \ref{DT_blk_diag}. 
    a) shows the root locus, where red corresponds to $\alpha > 0$ (the 0-deg root locus), and blue corresponds to $\alpha < 0$ (the 180-deg root locus).
    b) shows $\spr (p_\alpha)$ versus $\alpha$, where $\alpha_\rmn$ and $\alpha_\rmp$ are indicated.
    Note that $\alpha_\rmp \approx 0.6$, whereas the closed-loop system is asymptotically stable for all $\alpha\in(1.05,1.2)$.}
    \label{fig:Ex_spr_alpha}
\end{figure}

\begin{exam}\label{exam:spr_alpha}
Let $G(\bfq) = \frac{(\bfq - 0.05 \pm \jmath 0.88)(\bfq - 1)}{\bfq^2 (\bfq + 0.05 \pm \jmath 0.88)}
= \frac{(\bfq^2 - 0.1 \bfq + 0.7769)(\bfq - 1)}{\bfq^2 (\bfq^2 + 0.1 \bfq + 0.7769)}.$
Fig. \ref{fig:Ex_spr_alpha}a) shows the root locus for $\alpha >0$ and $\alpha <0$, and Fig. \ref{fig:Ex_spr_alpha}b) shows  $\spr(p_\alpha)$ versus $\alpha$, which indicates that there exists $\alpha > \alpha_\rmp$ such that $\spr(p_{\alpha}) < 1.$
\end{exam}

\section{State Space Analysis of the Linear Feedback System} \label{sec:SSfeedback}

Let $(A, B, C)$ be a minimal realization of $G$ with state $x_k \in \BBR^n$ at step $k$ so that $A$ is asymptotically stable.
The linear feedback system in Fig. \ref{DT_blk_diag} has the closed-loop dynamics
\begin{align}
    x_{k+1} & = (A + \alpha B C) x_k + \alpha B v_k, \label{xkEqLF}\\
    y_k &= C x_k.  \label{ykEqLF}
\end{align}
Note that, for all $\alpha\ne0,$ $(A + \alpha B C,\alpha B, C)$ is a minimal realization of $G_{yv,\alpha},$
and thus $p_\alpha = D-\alpha N = \chi_{A + \alpha B C}.$

\begin{lem}\label{CXiLem}
Let $\xi \in \BBC^n$ be an eigenvector of $A+\alpha BC.$ 
Then, $C \xi \ne 0.$
\end{lem}

{\bf Proof.}
Suppose that $C \xi =0.$
Since $\xi \in \BBC^n$ is an eigenvector of $A+\alpha BC,$ it follows that $\xi$ is an eigenvector of $A.$
Hence, for all $i \in \{0,1,\ldots,n-1\}$, $CA^i \xi =0.$
Since $(A,C)$ is observable, it follows that $\xi=0,$ which is a contradiction.
\hfill $\square$

The following lemma concerns the linear feedback system in Fig. \ref{DT_blk_diag} with $v\equiv0.$
In this case, \eqref{xkEqLF} and \eqref{ykEqLF} can be written as 
\begin{align}
    x_k &= (A+\alpha BC)^k x_0, \label{xkEqLF2}\\
    y_k &= C x_k.  \label{ykEqLF2}
\end{align}
Recall from Lemma \ref{RLasy} that, for all $|\alpha|$ sufficiently large, $A + \alpha BC$ has at least $n-m$ eigenvalues with absolute value greater than 1, all of which are simple.

\begin{lem}\label{propGyvalphaNS}
Consider the linear feedback system in Fig. \ref{DT_blk_diag} with $v\equiv0.$
Let $\alpha \in \BBR,$ 
assume there exists a simple eigenvalue $\lambda \in \spec(A+\alpha BC)$ such that $|\lambda| > 1$, 
let $\xi\in\BBC^n$ be an associated eigenvector,
let $\SX\subset\BBC^n$ be the $n-1$-dimensional subspace spanned by the eigenvectors and generalized eigenvectors associated with the remaining eigenvalues of $A+\alpha BC,$   
and assume that $x_0 \notin \SX.$ 
Then the following statements hold:
\begin{enumerate}
    \item  For all $k\ge0$, $x_k \notin \SX.$
    \item  $\limsup_{k \to \infty} |y_k| = \infty.$ 
\end{enumerate}
\end{lem}

 {\bf Proof.} 
Since ($A,B$) is controllable, Fact 7.15.10 of \cite[p. 599]{bernstein2018} implies that $A$ is cyclic.
Since, in addition, ($A+\alpha BC,B$) is controllable, it follows that $A+\alpha BC$ is cyclic.
Therefore, each eigenvalue of $A+\alpha BC$ has exactly one associated eigenvector.
Let $\lambda_1, \ldots, \lambda_{r}$ be the distinct eigenvalues of $A+\alpha BC,$ for all $j \in \{1, \ldots, r\},$ let $n_j$ be the algebraic multiplicity of $\lambda_j,$ and, for all $i \in \{1, \ldots, n_j\},$ let $\xi_{j,i}$ be a generalized eigenvector of $A+\alpha BC$ corresponding to $\lambda_j$ such that $(A+\alpha BC - \lambda_j I)^{i-1} \xi_{j,i} \ne0,$ $(A+\alpha BC - \lambda_j I)^{i} \xi_{j,i} =0,$ and $(\xi_{j, 1}, \ldots, \xi_{j, n_j})$ is a Jordan chain of $A+\alpha BC$ associated with $\lambda_j.$
Note that, for all $j\in \{1, \ldots, r\},$ $\xi_{j,1}$ is an eigenvector associated with $\lambda_j$.
Without loss of generality, define $\lambda_1 \isdef \lambda$ and $\xi_{1,1} \isdef \xi.$
Note that, since $\lambda$ is simple, it follows that $n_1 = 1.$
Next, it follows from the Jordan decomposition and equation (7.8.5) from \cite[p. 594]{meyer2000} that
\begin{equation}\label{Lem_ABCpow}
    (A + \alpha BC)^k 
    = S \matl  J_1^k & & \\ & \ddots & \\ & & J_r^k \matr  S^{-1},
\end{equation}
where $S\isdef [\xi_1 \ \cdots \ \xi_r]\in\BBC^{n\times n},$ for all $j\in\{1, \ldots, r\},$ $\xi_j \isdef [\xi_{j, 1} \ \cdots \ \xi_{j, n_j}]\in\BBC^{n\times n_j},$ $J_j \in \BBC^{n_j \times n_j}$ is the Jordan block associated with the eigenvalue $\lambda_j$ of $A + \alpha BC,$ and
\begin{equation}\label{Lem_Jpow}
    \resizebox{\columnwidth}{!}{%
    $J_j^k = \matl  \lambda_j^k & \binom{k}{1} \lambda_j^{k-1} & \binom{k}{2} \lambda_j^{k-2} & \cdots & \cdots & \binom{k}{n_j-1} \lambda_j^{k - n_j+1} \\
    & \lambda_j^k & \binom{k}{1} \lambda_j^{k-1} & \cdots & \cdots & \binom{k}{n_j-2} \lambda_j^{k - n_j+2}\\
    & & \ddots & \ddots & \vdots & \vdots\\
    & & & \ddots & \ddots & \vdots\\
    & & & & \lambda_j^k & \binom{k}{1} \lambda_j^{k-1} \\
    & & & & & \lambda_j^k\matr,$
    }
\end{equation}
where, for all $k < i,$ $\binom{k}{i}\isdef 0.$
Since $S$ is invertible, the $n$ linearly independent generalized eigenvectors 
$\xi,\xi_{2, 1}, \ldots, \xi_{2, n_2},\ldots,\xi_{r, 1}, \ldots, \xi_{r, n_r}$
comprise a basis of $\BBC^n$. 
Therefore, it follows that, for all $j\in \{1, \ldots, r\}$ and $i \in \{1, \ldots, n_j\},$ there exists $\beta_{j,i} \in \BBC$ such that
\begin{align}
   x_0 = \sum_{j=1}^{r} \sum_{i=1}^{n_j} \beta_{j,i} \xi_{j,i} 
    &=  \beta_{1,1} \xi  
    + \sum_{j=2}^{r}\sum_{i=1}^{n_j} \beta_{j,i} \xi_{j,i} \nn \\
    &= S \matl \beta_1^\rmT & \dots & \beta_r^\rmT \matr^\rmT,\label{Lem_x0Eq}
\end{align}
where 
$\beta_j \isdef [  \beta_{j, 1} \ \cdots \ \beta_{j, n_j}] ^\rmT\in\BBC^{n_j\times 1}.$
It thus follows from \eqref{xkEqLF2}, \eqref{Lem_ABCpow}, \eqref{Lem_Jpow}, and \eqref{Lem_x0Eq} that, for all $k\ge0,$
\begin{align}
    x_k
    &= S \matl  J_1^k & & \\ & \ddots & \\ & & J_r^k \matr   \matl \beta_1 \\ \vdots \\ \beta_r\matr  = S \matl J_1^k \beta_1 \\ \vdots \\ J_r^k \beta_r\matr \nn \\ 
    &= \sum_{j=1}^{r} \sum_{i=1}^{n_j} \gamma_{j,i}(k) \lambda_j^k \xi_{j,i},
\end{align}
where, for all $j\in\{1, \ldots, r\}$, $i\in\{1, \ldots, n_j\},$ and $k\ge0,$
\begin{equation}\label{Lem_gamma}
    \gamma_{j,i} (k) \isdef 
    \beta_{j, i} +  \sum_{l=2}^{n_j - i + 1} \binom{k}{l-1} \lambda_j^{1-l} \beta_{j, l + i - 1}.
\end{equation}
In particular,
\begin{equation}\label{Lem_gamma_11}
    \gamma_{1,1} (k)
    = \beta_{1,1}.
\end{equation}
Hence, 
\begin{align}
    x_k &=  \beta_{1,1}\lambda^k  \xi   
    + \sum_{j=2}^{r} \sum_{i=1}^{n_j}   \gamma_{j,i}(k) \lambda_j^k \xi_{j,i}, 
    %
    \label{Lem_xkeq}
\end{align}
and thus   it follows from \eqref{ykEqLF2} that  
\begin{equation}\label{Lem_ykeq2}
    y_k = p_1(k) \lambda^k   
    + \sum_{j=2}^{r}p_j (k) \lambda_j^k,
\end{equation}
where  
\begin{align}
    p_1(k)  &\isdef  \beta_{1,1}  C\xi,
\end{align}
and, for all $j=2,\ldots,r,$
\begin{equation}
\resizebox{\columnwidth}{!}{%
$p_j(k) \isdef \sum_{i=1}^{n_j} \left(\beta_{j,  i} +  \sum_{l=2}^{n_j - i + 1} \binom{k}{l-1}  \beta_{j, l + i - 1}\lambda_j^{1-l}\right) C\xi_{j,i}.$
}
\end{equation}
Note that $p_1(k),\ldots,p_r(k)$ are polynomials in $k$ with complex coefficients.
Now, the $n-1$-dimensional subspace   spanned by the eigenvectors and generalized eigenvectors associated with the eigenvalues $\lambda_2,\ldots,\lambda_r$ of $A+\alpha BC$ 
is given by
\begin{equation}
    \SX=\SR([\xi_{2, 1}\ \cdots \ \xi_{2, n_2}     \ \cdots \ \xi_{r, 1} \cdots \xi_{r, n_r}]).
\end{equation}
It follows from \eqref{Lem_x0Eq} that, since $x_0\notin\SX,$ $\beta_{1,1}\ne0,$  and thus it follows from \eqref{Lem_xkeq} that, for all $k\ge0,$ $x_k \notin \SX,$ which proves {\it i}). 
Next, Lemma \ref{CXiLem} implies that $C\xi \ne 0.$
Hence, $p_1$ is not the zero polynomial.
Finally, since $p_1$ is not the zero polynomial and $|\lambda| > 1,$  Proposition \ref{propA3} implies that $\limsup_{k\to\infty}\abs{y_k}=\infty,$ which proves {\it ii}).
\hfill $\square$

The following alternative characterization of $\SX$ is worth noting.

\begin{prop}
Let $\alpha\in\BBR,$ assume there exists a simple eigenvalue $\lambda \in \spec(A+\alpha BC)$ such that $|\lambda| > 1$, define $\SX$ as in Lemma \ref{propGyvalphaNS}, and define the polynomial $q(z)\isdef\chi_{A+\alpha BC}(z)/(z-\lambda).$
Then $\SX = \SN(q(A+\alpha BC)).$
\end{prop}

The following is a corollary of Lemma \ref{propGyvalphaNS}.

\begin{cor}\label{corGyvalphaNS}
Consider the linear feedback system in Fig. \ref{DT_blk_diag} with $v\equiv0.$
Let $\alpha \in \BBR,$ 
assume there exists a simple eigenvalue $\lambda \in \spec(A+\alpha BC)$ such that $|\lambda| > 1$, 
let $\xi\in\BBC^n$ be an associated eigenvector, 
let $\SX\subset\BBC^n$ be the $n-1$-dimensional subspace   spanned by the eigenvectors and generalized eigenvectors associated with the remaining eigenvalues of $A+\alpha BC$, 
let $k_0 \ge0,$ assume that $x_{k_0} \notin \SX,$  and let $M >|y_{k_0}|$.
Then, the following statements hold:
\begin{enumerate}
    \item  For all $k\ge k_0$, $x_k \notin \SX.$
    \item  There exists $k_1 > k_0$ such that the following statements hold:\\
    $a$)  For all $k\in \{k_0, \ldots, k_1 - 1\},$ $|y_{k}| < M$.\\
    $b$)  $|y_{k_1}| \ge M.$
\end{enumerate}
%
%
%
%
%
\end{cor}


\section{Analysis of the Lur'e System}\label{sec:lure}

\begin{figure}[h!]
    \centering
     \resizebox{0.5\columnwidth}{!}{%
    \begin{tikzpicture}[>={stealth'}, line width = 0.25mm]
    \node [input, name=input] {};
    \node [sum, right = 0.5cm of input] (sum1) {};
    \node[draw = white] at (sum1.center) {$+$};
    \node [smallblock,  rounded corners, right = 0.4cm of sum1, minimum height = 0.6cm, minimum width = 0.8cm] (system) {$\alpha  G(\bfq)$};
    \node [smallblock, rounded corners, below = 0.25cm of system, minimum height = 0.6cm, minimum width = 0.8cm] (satF) {${\rm sat}_1$};
    \draw [->] (input) -- node[name=usys, xshift = -0.2cm, yshift = 0.2cm] {\scriptsize$v \equiv 0$} (sum1.west);
    \draw[->] (sum1.east) -- node [near end, above, xshift = -0.12cm] {\scriptsize$\nu$} (system.west);
    \node [output, right = 0.5cm of system] (output) {};
    \draw [->] (system) -- node [name=y,near end, xshift = -0.15cm]{} node [near end, above, xshift = 0.1cm] {\scriptsize$y$}(output);
    \draw [->] (y.center) |- (satF.east);
    \draw (satF.west) -| (sum1.south);
    \end{tikzpicture}
    }
    \caption{\footnotesize Discrete-time Lur'e system.}
    \label{DTL_wSat_blk_diag}
\end{figure}
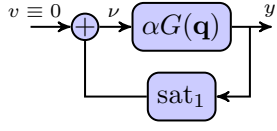

We now consider the discrete-time Lur'e system in Fig. \ref{DTL_wSat_blk_diag}, which has the closed-loop dynamics
\begin{align}
    x_{k+1} & = A x_k + \alpha B {\rm sat}_1 (y_k), \label{xkEqNL}\\
    y_k &= C x_k. \label{ykEqNL}
\end{align}

\indent\begin{prop}\label{DTL_prop_ybd}
Let $y$ be the output of the discrete-time Lur'e system in Fig. \ref{DTL_wSat_blk_diag}. Then, $y$ is bounded.
\end{prop}

{\bf Proof:} Since $G$ is asymptotically stable, it is bounded-input, bounded-output stable.
Since the input of $G$ is bounded, it follows that $y$ is bounded.
\hfill $\square$

\begin{figure}[h!]
    \centering
     \resizebox{\columnwidth}{!}{%
    \begin{tikzpicture}[>={stealth'}, line width = 0.25mm]
    
    \node [input, name=input1] {};
    \node [smallblock, fill=green!20, rounded corners, below right = -0.5cm and -0.45cm of input1, minimum height = 1.25cm, minimum width = 3.75cm] (stg1) {};
    \node[draw = none] at ([xshift = -0.14cm, yshift = -0.5cm]input1.center) {\scriptsize $\SSS_2$};
    \node [sum, right = 0.55cm of input1] (sum1) {};
    \node[draw = none] at (sum1.center) {$+$};
    \node [smallblock, rounded corners, right = 0.4cm of sum1, minimum height = 0.6cm, minimum width = 0.8cm] (system1) {$\alpha  G(\bfq)$};
    \draw [->] (input1) -- node[name=usys, xshift = -0.2cm, yshift = 0.2cm] {\scriptsize$v = 0$} (sum1.west);
    \draw[->] (sum1.east) -- node [near end, above, xshift = -0.12cm] {\scriptsize$\nu$} ([xshift=0.07cm]system.west);
    \node [output, right = 0.5cm of system1] (output1) {};
    \draw [->] (system1) -- node [name=y,near end, xshift = -0.15cm]{} node [near end, above, xshift = 0.1cm] {\scriptsize$y$}(output1);
    \node [coordinate, below = 0.25cm of system1] (midpoint1) {};
    \draw [->] (y.center) |- (midpoint1) -| (sum1.south);
    
     \node [input, above left = 2 cm and 2 cm](input2){};
     \node [smallblock, fill=green!20, rounded corners, below right = -0.5cm and -0.345cm of input2, minimum height = 0.95cm, minimum width = 2.9cm] (stg2) {};
     \node[draw = none] at ([yshift = -0.2cm]input2.center) {\scriptsize $\SSS_1$};
     \node [smallblock,  rounded corners, right = 0.75cm of input2, minimum height = 0.6cm, minimum width = 0.8cm] (system2) {$\alpha  G(\bfq)$};
     \node [output, right = 0.35cm of system2] (output2) {};
     \draw [->] (input2.center) -- node[xshift = -0.2cm, yshift = 0.2cm] {\scriptsize$\nu = 1$} (system2.west);
     \draw [->] (system2.east) -- node [near end, xshift = -0.25cm]{} node [near end, above, xshift = 0.1cm] {\scriptsize$y$}(output2);
     
     \node [input, above right = 2 cm and 2.5 cm](input3){};
     \node [smallblock, fill=green!20, rounded corners, below right = -0.5cm and -0.44cm of input3, minimum height = 0.95cm, minimum width = 3cm] (stg3) {};
     \node[draw = none] at ([xshift = -0.12cm, yshift = -0.2cm]input3.center) {\scriptsize $\SSS_3$};
     \node [smallblock,  rounded corners, right = 0.75cm of input3, minimum height = 0.6cm, minimum width = 0.8cm] (system3) {$\alpha  G(\bfq)$};
     \node [output, right = 0.35cm of system3] (output3) {};
     \draw [->] (input3.center) -- node[xshift = -0.2cm, yshift = 0.2cm] {\scriptsize$\nu = -1$} (system3.west);
     \draw [->] (system3.east) -- node [near end, xshift = -0.25cm]{} node [near end, above, xshift = 0.1cm] {\scriptsize$y$}(output3);
     
     \draw [->,red] ([xshift = -1 cm]stg1.north) -- node[xshift = -0.35cm, yshift = -0.2cm] {\scriptsize$y_k \geq 1$} (stg2.south);
     \draw [<-,red] ([xshift = -0.5 cm]stg1.north) -- node[xshift = 0.55cm, yshift = 0.2cm] {} ([xshift = 0.5 cm]stg2.south);
     \draw[->,red] ([yshift = 0.15cm]stg1.west) arc (60:300:0.2cm);
     \node[draw = none,red] at ([xshift = -1.1cm,yshift = -0.05cm]stg1.west) {\scriptsize $y_k \in (-1, 1)$};

     \draw [->,red] ([xshift = 1 cm]stg1.north) -- node[xshift = 0.5cm, yshift = -0.2cm] {\scriptsize$y_k \leq -1$} (stg3.south);
     \draw [<-,red] ([xshift = 0.5 cm]stg1.north) -- node[xshift = -1.12cm, yshift = -0.05cm] {\scriptsize$y_k \in (-1, 1)$} ([xshift = -0.5 cm]stg3.south);

     \draw [->,red] ([yshift = 0.1 cm]stg2.east) -- node[xshift = -0.05cm, yshift = -0.375cm] {\scriptsize$y_k \geq 1$} ([yshift = 0.1 cm]stg3.west);
     \draw [<-,red] ([yshift = -0.1 cm]stg2.east) -- node[xshift = -0.05cm, yshift = 0.375cm] {\scriptsize$y_k \leq -1$} ([yshift = -0.1 cm]stg3.west);
     
     \draw[->,red] ([yshift = 0.2cm]stg2.west) arc (60:300:0.2cm);
     \node[draw = none,red] at ([xshift = -0.82cm]stg2.west) {\scriptsize $y_k \geq 1$};
     
     \draw[->,red] ([yshift = -0.2cm]stg3.east) arc (60:300:-0.2cm);
     \node[draw = none,red] at ([xshift = 0.88cm]stg3.east) {\scriptsize $y_k \leq -1$};
    
    \end{tikzpicture}
    }
    \caption{\footnotesize Transition diagram  for the discrete-time Lur'e system.}
    \label{DTL_Swicthed_blk_diag}
\end{figure}
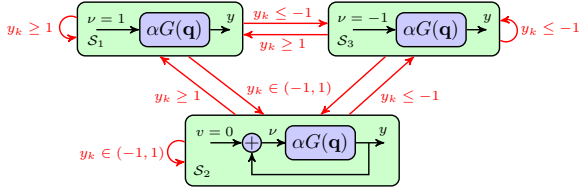

Let $S_k$ denote the dynamics of the discrete-time Lur'e system at step $k\geq 0,$ and consider the transition diagram in Fig. \ref{DTL_Swicthed_blk_diag}, where, 
%
%
%
%
%
%
for all $k\ge 0$ such that $y_k \ge 1,$ $S_k = \SSS_1$, and thus $x_{k+1} = Ax_k + \alpha B;$ 
for all $k\ge 0$ such that $|y_k|<1,$ $S_k = \SSS_2$, and thus $x_{k+1} = (A + \alpha BC) x_k;$
and,
for all $k\ge 0$ such that $y_k \le -1,$ $S_k = \SSS_3$, and thus $x_{k+1} = Ax_k - \alpha B.$ 

 Henceforth, the terminology $\lim_{k\to\infty} y_k$ {\it does not exist} means that there does not exist a real number $\overline{y}$ such that $\lim_{k\to\infty} y_k=\overline{y}.$
For example, $\lim_{k\to\infty}(-1)^k$, $\lim_{k\to\infty}k,$ and $\lim_{k\to\infty}(-1)^kk$ do not exist.
The following lemma is needed.

\begin{lem}\label{lemma1notinspec}
Let $\alpha\in\BBR.$  Then the following statements hold:
\begin{enumerate}
    \item $1\notin \spec(A+\alpha BC)$.
    \item  If $\lim_{k\to\infty} (A+\alpha BC)^k$ exists, then the limit is zero.
    \item  If, for $y_k$ given by \eqref{xkEqLF} and \eqref{ykEqLF} with $v_k \equiv 0,$ $\lim_{k\to\infty} y_k$ exists, then the limit is zero.
\end{enumerate} 
\end{lem}

{\bf Proof:}  
To prove {\it i}), first consider the case where $\alpha=0.$   Since $A$ is asymptotically stable, it follows that $1\notin \spec(A)$.
Now, let $\alpha\ne0.$ Then $(A + \alpha B C,\alpha B, C)$ is a minimal realization of $G_{yv,\alpha},$
and thus $\chi_{A + \alpha B C} = p_\alpha = D-\alpha N.$
Since $N$ and $D$ are coprime and $G(1)=0$, it follows that $N(1) = 0$ and $D(1)\ne0.$
Hence, $\chi_{A + \alpha B C}(1) = D(1)-\alpha N(1) = D(1) \ne 0.$
Therefore, $1\notin \spec(A+\alpha BC)$.

 {\it ii}) and {\it iii}) follow from Lemma \ref{lemmalimits}.
\hfill$\square$

%
%

%

\begin{lem}\label{lemmaDTTDL3}
Let $y$ be the output of the discrete-time Lur'e system in Fig. \ref{DTL_wSat_blk_diag} with $\alpha \in \BBR \backslash\{0\}$.
Consider the following statements:\\
    $a$) $\lim_{k \to \infty} y_k$ exists, and $\lim_{k \to \infty} y_k=0.$\\
    $b$) $\lim_{k \to \infty} y_k$ exists, and $\lim_{k \to \infty} y_k\in(-1,1).$\\
    $c$) $\lim_{k \to \infty} y_k$ exists, and $\lim_{k \to \infty} y_k\in[-1,1].$\\
    $d$) $\lim_{k \to \infty} y_k$ exists.\\
    Then $a$) $\Longleftrightarrow$ $b$) $\Longleftrightarrow$ $c$) $\Longleftrightarrow$ $d$).
\end{lem}

 {\bf Proof:} Note that $a$) $\Longrightarrow$ $b$) $\Longrightarrow$ $c$) $\Longrightarrow$ $d$).
%
%

 To prove $b)$ $\Longrightarrow$ $a),$ 
note that $\lim_{k \to \infty} y_k\in(-1,1)$ implies that there exists $k_1 \ge n$ such that, for all $k\ge k_1$, $|y_k|<1$ and $S_k = \SSS_2.$
Hence, for all $k\ge k_1$ $y_k$ is given by \eqref{xkEqLF2} and \eqref{ykEqLF2} with $x_0$ replaced by $x_{k_1}$.
Since $\lim_{k \to \infty} y_k = \lim_{k \to \infty} C(A+\alpha BC)^{k - k_1} x_{k_1}$ exists, Lemma \ref{lemma1notinspec} implies that $\lim_{k \to \infty} y_k = 0.$
%
%

To prove $c)$ $\Longrightarrow$ $b),$ consider the case where $\lim_{k \to \infty} y_k = 1,$ so that $\lim_{k \to \infty} \nu_k = 1.$
Next, note that it follows from input-to-state stability for linear time-invariant discrete-time systems  (see \cite[Example 3.4]{ISSCT}) that
\begin{equation}
    \lim_{k\to\infty} \sum_{i=0}^{k-1} CA^{k-1-i} B(\nu_{i}-1) = 0.\label{ISS}
\end{equation}
Since $\lim_{k\to\infty} CA^{k}x_{0} = 0$ and $C(I - A)^{-1} B = G(1) = 0,$ it follows from \eqref{ISS} that
\small
\begin{align*}
    1 & = \lim_{k\to\infty} y_k \\
    &=\lim_{k\to\infty}CA^{k} x_{0} 
      + \alpha\lim_{k\to\infty}\sum_{i=0}^{k-1} CA^{k-1-i} B\nu_i\\
    &=\alpha\lim_{k\to\infty}\sum_{i=0}^{k-1} CA^{k-1-i} B \\
    & \quad+\alpha\lim_{k\to\infty}\sum_{i=0}^{k-1} CA^{k-1-i} B(\nu_i-1)\\
    %
    %
    %
    &=\alpha \sum_{i=0}^{\infty} CA^{i} B 
    %
    %
    %
     =\alpha \ C(I-A)^{-1} B\\ 
    & = 0,
\end{align*}
\normalsize
which is a contradiction.
Hence, $\lim_{k \to \infty} y_k < 1.$
A similar argument implies that $\lim_{k \to \infty} y_k > -1.$ 
Hence, $\lim_{k \to \infty} y_k\in(-1,1).$

 To prove $d$) $\Longrightarrow$ $c$), suppose that $\lim_{k \to \infty} y_k>1.$
Then there exists $k_1 \ge n$ such that, for all $k\ge k_1$, $y_k\ge1,$ and thus $S_k = \SSS_1.$  
Hence, for all $k\ge k_1,$ $y_k$ is given by 
\begin{align*}
    y_k &= CA^{k-k_1} x_{k_1} + \alpha\sum_{i=0}^{k-k_1-1} CA^{k-k_1-1-i} B \\
    &= CA^{k-k_1} x_{k_1} + \alpha\sum_{i=0}^{k-k_1-1} CA^i B.
\end{align*}
Since $\lim_{k\to\infty} A^k = 0$ and $C(I - A)^{-1} B = G(1) = 0,$ it follows that
\begin{align*}
    \lim_{k\to\infty} y_k &= \lim_{k\to\infty} CA^{k-k_1} x_{k_1} + \lim_{k\to\infty} \alpha\sum_{i=0}^{k-k_1-1} CA^i B \\
    &=\alpha \sum_{i=0}^{\infty} CA^i B = \alpha C (I - A)^{-1}B = 0.
\end{align*}
%
%
%
%
Hence, there exists $k_2>k_1$ such that $y_{k_2}<1,$ which is a contradiction.
Therefore, $\lim_{k \to \infty} y_k\le1.$
Similarly, $\lim_{k \to \infty} y_k\ge-1.$ 
Hence, $\lim_{k \to \infty} y_k\in[-1,1].$ 
\hfill$\square$

\begin{lem}\label{lemmaDTTDL1}
Consider the discrete-time Lur'e system in Fig. \ref{DTL_Swicthed_blk_diag}.
Let $\alpha \in \BBR,$
assume there exists a simple eigenvalue $\lambda \in \spec(A+\alpha BC)$ such that $|\lambda| > 1$, let $\xi\in\BBC^n$ be an associated eigenvector,
let $\SX\subset\BBC^n$ be the $n-1$-dimensional subspace spanned by the eigenvectors and generalized eigenvectors associated with the remaining eigenvalues of $A+\alpha BC,$ 
let $k_0 \ge0$, and assume that $x_{k_0} \notin \SX$ and $S_{k_0} = \SSS_2$.
Then there exists $k_1 > k_0$ such that, for all $k \in \{k_0, k_0 + 1, \ldots, k_1-1\},$ $S_{k} = \SSS_2$ and $x_{k}\notin\SX,$  and such that 
$x_{k_1} \notin \SX$ and $S_{k_1} \ne \SSS_2.$
\end{lem}

{\bf Proof:}
For all $k\ge k_0$ such that $S_k = \SSS_2,$ the system dynamics are given by the discrete-time linear feedback system in Fig. \ref{DT_blk_diag} with $v\equiv0.$
Hence, the result follows from Corollary \ref{corGyvalphaNS} with $M = 1.$
\hfill $\square$

The following result gives sufficient conditions under which the response of the discrete-time Lur'e system is bounded and nonconvergent.

\begin{theo}\label{LureTheo}
Consider the discrete-time Lur'e system in Fig. \ref{DTL_Swicthed_blk_diag}.
Let $\alpha \in \BBR,$
assume there exists a simple eigenvalue $\lambda \in \spec(A+\alpha BC)$ such that $|\lambda| > 1$, 
let $\xi\in\BBC^n$ be an associated eigenvector, and
let $\SX\subset\BBC^n$ be the $n-1$-dimensional subspace   spanned by the eigenvectors and generalized eigenvectors associated with the remaining eigenvalues of $A+\alpha BC$.
Furthermore, assume that, if $S_0 = \SSS_2,$ then $x_{0} \notin \SX$, and, for all $k\ge 0$ such that $S_k \ne \SSS_2$ and $S_{k+1} = \SSS_2$, it follows that $x_{k+1} \notin \SX.$
Then $y$ is bounded and $\lim_{k\to \infty} y_k$ does not exist.
\end{theo}

{\bf Proof:} Proposition \ref{DTL_prop_ybd} implies that $y$ is bounded.
Suppose that $\lim_{k \to \infty} y_k$ exists, and thus Lemma \ref{lemmaDTTDL3} implies that $\lim_{k \to \infty} y_k=0.$ 
Now, let $k_0\ge0$ be the smallest nonnegative integer such that, for all $k\ge k_0,$ $|y_k| < 1$ and thus $S_k = \SSS_2.$  
In the case where $k_0 = 0,$ it follows that $x_0\notin\SX,$ and thus Lemma \ref{lemmaDTTDL1} implies that there exists $k>k_0$ such that $|y_k|>1.$
Alternatively, in the case where $k_0>0$, note that $\SSS_{k_0-1}\ne\SSS_2$ and $x_{k_0}\notin\SX,$ and thus 
Lemma \ref{lemmaDTTDL1} implies that there exists $k>k_0$ such that $|y_k|>1.$ 
In both cases, the existence of $k>k_0$ such that $|y_k|>1$ contradicts the fact that, for all $k \ge k_0,$ $|y_k| < 1.$
Therefore, $\lim_{k \to \infty} y_k$ does not exist.
\hfill$\square$

The following result shows that, for almost all initial conditions $x_0,$ the hypotheses of Theorem \ref{LureTheo} are satisfied.

\begin{theo}\label{LureTheo2}
Consider the discrete-time Lur'e system in Fig. \ref{DTL_Swicthed_blk_diag}.
Let $\alpha \in \BBR,$
assume that $A$ is nonsingular,
assume there exists a simple eigenvalue $\lambda \in \spec(A+\alpha BC)$ such that $|\lambda| > 1$, 
let $\xi\in\BBC^n$ be an associated eigenvector, and
let $\SX\subset\BBC^n$ be the $n-1$-dimensional subspace spanned by the eigenvectors and generalized eigenvectors associated with the remaining eigenvalues of $A+\alpha BC$.
Then, for almost all $x_0\in\BBR^n,$
$y$ is bounded and $\lim_{k\to \infty} y_k$ does not exist.
\end{theo}

{\bf Proof:}
Define $f\colon\BBR^n\to\BBR^n$ by $f(x)= A x+\alpha B {\rm sat}(Cx).$
%
%
%
%
Letting $\SE$ denote a proper affine subspace of $\BBR^n,$ it follows that 
\begin{align*}
    f^{-1} (\SE) &\subset A^{-1} (\SE - \alpha B) \cup  A^{-1} (\SE + \alpha B)\\
    &\quad\cup (A + \alpha BC)^{-1} (\SE).  
\end{align*}
Hence, the inverse image of every subset of the union of a finite number of proper affine subspaces of $\BBR^n$ is a subset of the union of a finite number of proper affine subspaces.
In particular, $f^{-1} (\SX)$ has measure zero.
Now, for all $k\ge1,$ define $f^{-k-1} (\SX)\isdef  f^{-1}(f^{-k} (\SX)).$ 
By induction, it follows that, for all $k\ge0,$ $ f^{-k}(\SX)$ a subset of the union of a finite number of proper affine subspaces and thus has measure zero.
Hence, $\cup_{k\ge1} f^{-k}(\SX)$ is a countable union of sets with measure zero, and thus has measure zero.
Therefore, for all $x_0\notin\cup_{k\ge1} f^{-k}(\SX),$ it follows that, for all $k\ge0,$  $x_k\notin \SX.$ %
\hfill$\square$


%
%
%

\section{Numerical Example}\label{sec:numericalEx}

A key assumption in Theorem \ref{LureTheo} is the requirement that, if, at step $k,$ the system changes either from $\SSS_1$ to $\SSS_2$ or from $\SSS_3$ to $\SSS_2,$ then the state $x_{k+1}$ is not contained in the $n-1$-dimensional subspace $\SX.$
Under this assumption, the state of the closed-loop system includes a component from the unstable subspace of the closed-loop dynamics.  Consequently, the norm of the state ultimately increases, and thus the system eventually leaves $\SSS_2.$
Theorem \eqref{LureTheo2} implies that this assumption is satisfied for almost all initial conditions, as demonstrated by the following example, where the response converges and thus is not self-excited.
However, under a small perturbation of the initial condition, the response is oscillatory, which illustrates the generic nature of the assumption on the initial condition.

\begin{exam}\label{exam:ProjConv}
Let $G(\bfq) =  \tfrac{\bfq - 1}{\bfq^2 - \bfq + 0.5},$ so that $\Theta_\rmn = \{\pi\}$ rad, $\Theta_\rmp = \{\pm {\rm acos} (3/4)\}$ rad, $\alpha_\rmn = -1.25,$ and $\alpha_\rmp = 0.5.$
The root locus of the closed-loop linear system is shown in Fig. \ref{fig:Ex_proj_rlocus_spr}, along with $\spr$ for a range of $\alpha.$
Note that, for all $\alpha > \alpha_\rmp,$ both eigenvalues of $A+\alpha BC$ are unstable, whereas, for all $\alpha < \alpha_\rmn,$ one eigenvalue of $A+\alpha BC$ is unstable and one eigenvalue of $A+\alpha BC$ is asymptotically stable.
\begin{figure}[h!]
    \centering
    \includegraphics[width = \columnwidth]{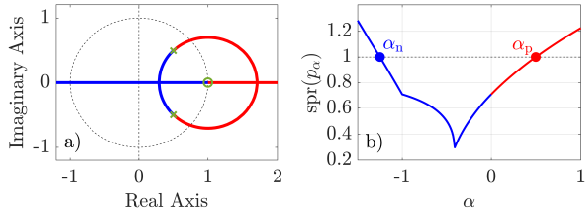}
    \caption{Example \ref{exam:ProjConv}. Root locus and $\spr(p_\alpha)$ for the closed-loop linear system. 
    a) shows the root locus, where red corresponds to $\alpha > 0$ (the 0-deg root locus), and blue corresponds to $\alpha < 0$ (the 180-deg root locus).
    b) shows $\spr (p_\alpha)$ versus $\alpha$, where $\alpha_\rmn$ and $\alpha_\rmp$ are indicated.}
    \label{fig:Ex_proj_rlocus_spr}
\end{figure}

Let the minimal realization of $G$ be given by
\begin{equation*}
    A = \begin{bmatrix} 1 & -0.5 \\ 1 & 0 \end{bmatrix}, \ B = \begin{bmatrix} 1 \\ 0 \end{bmatrix}, \ C = \begin{bmatrix} 1 & -1 \end{bmatrix},
\end{equation*}
and let $\alpha = -2.5 < \alpha_\rmn.$
Then $\spec (A + \alpha BC) = \{\lambda, \lambda_2\},$ where $\lambda = - 0.75 - 0.25\sqrt{41}\approx -2.35$ and 
$\lambda_2 = - 0.75 + 0.25\sqrt{41}\approx 0.85.$
Hence,  $|\lambda| > 1$ and $|\lambda_2| < 1.$ 
Furthermore, let $\xi = [ -0.75 - 0.25 \sqrt{41} \ \ \ 1  ]^\rmT$ and $\xi_2 = [ -0.75 + 0.25 \sqrt{41} \ \ \ 1 ]^\rmT$ be eigenvectors of $A + \alpha BC$ associated with $\lambda$ and $\lambda_2$, respectively.
Hence, $\SX=\{\alpha\xi_2\colon \alpha\in\BBR\}.$
Now, let $\Psi \in \BBC^2$ satisfy $||\Psi|| = 1$ and $\SX = \{\Psi\}^{\perp},$ that is, $\Psi^*\xi_2 = 0,$ and define the projector $P \isdef \Psi \Psi^{*}.$
Note that, for all $x \in \BBR^2,$ $x \notin \SX$ if and only if $||P x|| \ne 0.$
Let $x_0$ be given by 
\begin{equation}\label{eq:x0C1}
    x_0 = [ -5.5-6.5 \sqrt{41} \ \ \  -51 ]^\rmT,
\end{equation}
such that $y_0 = C x_0 \approx 3.88 > 1.$ 
Then, it follows from the system dynamics with exact symbolic computation that
\footnotesize
\begin{align*}
    x_1 &= A x_0 + \alpha B =  [ 17.5 - 6.5 \sqrt{41} \ \ \ \ -5.5 -6.5 \sqrt{41}  ]^\rmT, \\
    x_2 &= A x_1 + \alpha B = [ 17.75 - 3.25 \sqrt{41} \ \ \ \  17.5 -6.5 \sqrt{41} ]^\rmT, \\ 
    x_3 &= A x_2 + \alpha B = [ 6.5 \ \ \ \ 17.75 -3.25 \sqrt{41} ]^\rmT, \\
    x_4 &= A x_3 + \alpha B = [ -4.875 + 1.625 \sqrt{41} \ \ \ \ 6.5 ]^\rmT.
\end{align*}
\normalsize
Hence, $y_1 = C x_1 = 23,$  $y_2 = C x_2 = 0.25 + 3.25 \sqrt{41} \approx 21.06 > 1,$  $y_3 = C x_3 = -11.25 + 3.25 \sqrt{41} \approx 9.56 > 1,$  $y_4 = C x_4 = -11.375 + 1.625 \sqrt{41} \approx -0.97  \in (-1, 1),$ and thus $S_0 = S_1 = S_2 = S_3 = \SSS_1,$ and $S_4 = \SSS_2.$
Note that $x_4 = 6.5 \xi_2.$
Hence, for all $k \ge 4,$
\begin{align}
    C (A + \alpha BC)^{k-4} x_4 &= 6.5 C (A + \alpha BC)^{k-4} \xi_2 \nn \\
    &= 6.5 \lambda_2^{k-4} C \xi_2. \label{ex:eq1}
\end{align}
Since $|C x_4| = 6.5 |C \xi_2| = |- 11.375 +  1.6250 \sqrt{41}| \approx 0.97 < 1,$ it follows from \eqref{ex:eq1} that, for all $k \ge 4,$
\small
\begin{align*}
     |C &(A + \alpha BC)^{k-4} x_4| = 6.5 |\lambda_2|^{k-4} |C \xi_2| \\ 
     &= |- 11.375 +  1.6250 \sqrt{41}| |-0.75 - 0.25 \sqrt{41}|^{k-4} \\
     &\approx 0.97 \ (0.85)^{k-4} < 1,
\end{align*}
\normalsize
and thus $S_k = \SSS_2.$
Hence,  
\begin{align*}
    \lim_{k\to \infty} y_k &=  \lim_{k\to \infty}  C (A + \alpha BC)^{k-4} x_4 \\
    &= 6.5 \ C \xi_2 \lim_{k\to \infty} \lambda_2^{k-4} = 0,
\end{align*}
and, for all $k \ge 4,$
\begin{align*}
    ||P x_k|| &= ||P (A + \alpha BC)^{k-4} x_4|| \\ 
                &= 6.5 \ ||\lambda_2||^{k-4} ||P \xi_2|| = 0.
\end{align*}
Therefore, 
$||P x_4|| = 0,$ which implies that $x_4 \in \SX,$ and thus the assumptions of Theorem \ref{LureTheo} are not satisfied.  
In this case, the output converges and thus the system does not have self-excited oscillations, as shown in Fig. \ref{fig:Ex_proj_conv}.

Next, let $x_0$ be given by
\begin{equation}\label{eq:x0C2}
    x_0 = [ -5.5-6.5 \sqrt{41} \ \ \  -51 + \varepsilon ]^\rmT,
\end{equation}
where $\varepsilon \isdef 10^{-12},$
which represents a small perturbation of \eqref{eq:x0C1}, such that $y_0 = C x_0 \approx 3.88 > 1.$
With the initial condition \eqref{eq:x0C2}, it follows that
\footnotesize
\begin{align*}
    x_1 &=  [ 17.5 - 6.5 \sqrt{41} -0.5 \varepsilon \ \ \ \ -5.5 -6.5 \sqrt{41}  ]^\rmT, \\
    x_2 &= [ 17.75 - 3.25 \sqrt{41} -0.5 \varepsilon \\ 
    & \hspace{8em} 17.5 -6.5 \sqrt{41} -0.5 \varepsilon ]^\rmT, \\ 
    x_3 &= [ 6.5 -0.25 \varepsilon \ \ \ \ 17.75 -3.25 \sqrt{41} -0.5 \varepsilon ]^\rmT, \\
    x_4 &= [ -4.875 + 1.625 \sqrt{41} \ \ \ \ 6.5 - 0.25 \varepsilon ]^\rmT.
\end{align*}
\normalsize
Hence, $y_1 = C x_1 = 23 -0.5 \varepsilon,$  $y_2 = C x_2 = 0.25 + 3.25 \sqrt{41} \approx 21.06 > 1,$  $y_3 = C x_3 = -11.25 + 3.25 \sqrt{41} + 0.25 \varepsilon \approx 9.56 > 1,$  $y_4 = C x_4 = -11.375 + 1.625 \sqrt{41} + 0.25 \varepsilon \approx -0.97  \in (-1, 1),$ and thus $S_0 = S_1 = S_2 = S_3 = \SSS_1,$ and $S_4 = \SSS_2.$
Defining $\kappa_1 \isdef 0.5 - 3 \sqrt{41}/82$ and $\kappa_2 \isdef 0.5 + 3 \sqrt{41}/82,$ it follows that
\begin{align*}
    \kappa_1 \xi + \kappa_2 \xi_2 &= \begin{bmatrix} -\sqrt{41}/8 + 9 \sqrt{41}/328 \\ 0.5 - 3 \sqrt{41}/82 \end{bmatrix} \\ 
    &+ \begin{bmatrix} \sqrt{41}/8 - 9 \sqrt{41}/328 \\ 0.5 + 3 \sqrt{41}/82  \end{bmatrix} = \begin{bmatrix} 0 \\ 1 \end{bmatrix}.
\end{align*}
Then, $x_4 = 6.5 \xi_2 + [ 0 \ \ \ -0.25 \varepsilon ] = (- 0.25 \kappa_2 \varepsilon) \xi + (6.5 - 0.25 \kappa_2 \varepsilon ) \xi_2 = \eta_1 \xi + \eta_2 \xi_2,$ where $\eta_1, \eta_2 \ne 0.$
Hence, for all $k \ge 4,$
\begin{align}
    C (A + \alpha BC)^{k-4} x_4 &= \eta_1 C (A + \alpha BC)^{k-4} \xi \nn \\
    &+ \eta_2 C (A + \alpha BC)^{k-4} \xi_2 \nn \\
    &= \eta_1 C \xi \lambda^{k-4} + \eta_2 C \xi_2 \lambda_2^{k-4}. \label{ex:eq2}
\end{align}
Since $\eta_1 C \xi \ne 0,$ $\eta_2 C \xi_2 \ne 0,$ $|\lambda| > |\lambda_2|,$ and $|\lambda| > 1,$ it follows from \eqref{ex:eq2} and Proposition \ref{propA3} that
\begin{align*}
    \limsup_{k\to\infty}&\abs{C (A + \alpha BC)^{k-4} x_4}  \\
    &= \limsup_{k\to\infty}\abs{\eta_1 C \xi \lambda^{k-4} + \eta_2 C \xi_2 \lambda_2^{k-4}} = \infty.
\end{align*}
Hence, there exists $k > 4$ such that $S_k \ne \SSS_2.$
Furthermore, note that
\begin{equation*}
    ||P x_4|| = ||\eta_1 P \xi + \eta_2 P \xi_2|| = ||\eta_1 P \xi|| \ne 0,
\end{equation*}
which implies that $x_4 \notin \SX.$
Therefore, with $x_0$ given by \eqref{eq:x0C2}, $x_4 \notin \SX,$ and thus the assumptions of Theorem \ref{LureTheo} are satisfied.  
In this case, Fig. \ref{fig:Ex_proj_conv} shows that the output does not converge and is bounded, and thus the system has self-excited oscillations. 
In fact, the asymptotic response is periodic with period 2 steps.

\begin{figure}[h]
    \centering
    \includegraphics[width = \columnwidth]{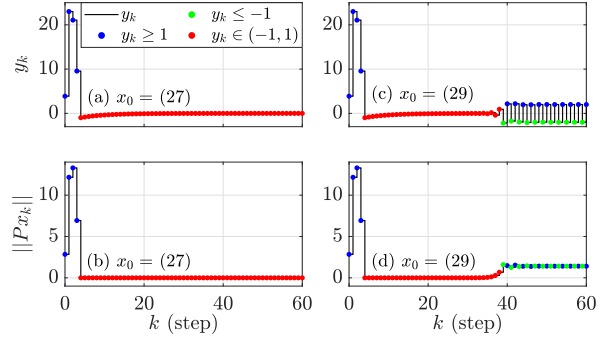}
    \caption{Example \ref{exam:ProjConv}:
    a) and b) show the response of the Lur'e system in the case where $x_0$ is given by \eqref{eq:x0C1} and c) and d) show the response of the Lur'e system in the case where $x_0$ is given by \eqref{eq:x0C2}. a) and c) show the system response $y_k$ versus $k.$ b) and d) show the norm $||P x_k||$ of the projection of $x_k$ onto $\SR(\Psi)$  versus $k.$}
    \label{fig:Ex_proj_conv}
\end{figure}
\end{exam}

\section{Conclusions and Future Work}\label{sec:conclusions}

This paper analyzed a discrete-time Lur'e system that exhibits self-excited oscillations.
This system involves an asymptotically stable linear system with a  zero at 1 connected in feedback with a piecewise-linear saturation  nonlinearity.
It was shown that, for sufficiently large loop gains, the response of the system is bounded and does not converge, and thus the system has self-excited oscillations.
A numerical example illustrated the conditions under which this discrete-time Lur'e system yields a self-excited response.
Future work will extend the Lur'e model to sigmoidal nonlinearities as well as the use of this model  for system identification.

\begin{appendices}

\section{}

\begin{lem} \label{appLem1}
Let $z_1,\ldots,z_m\in\BBC$ be distinct, assume that, for all $i=1,\ldots,m,$ $|z_i|\ge1,$ let $a_1,\ldots,a_m\in\BBC,$ and assume that  
$\lim_{k\to\infty}\sum_{i=1}^m a_i z_i^k=0.$ 
Then, $a_1=\cdots=a_m=0.$
\end{lem}

{\bf Proof: }
First, consider the case where $m=1.$
Suppose that $a_1\ne0.$  Then, $\lim_{k\to\infty}a_1z_1^k=0,$ and thus $\lim_{k\to\infty}|z_1|^k=0,$ which contradicts the assumption that $|z_1|\ge1.$
Therefore, $a_1 = 0.$

Next, consider the case where $m\ge2.$
Let $s\in\{1,\ldots,m\},$ and define the polynomial
\begin{equation*}
     p_s(z)\isdef \prod_{ \atopJ{1\le t\le m}{t\ne s}}\frac{z-z_t}{z_s-z_t}.
\end{equation*}
Writing $p_s(z) = \sum_{j=0}^{m-1}b_j z^j,$ where $b_0,\ldots,b_{m-1} \in \BBC$,  it follows that, for all $k\ge1,$
\begin{equation}
	\sum_{i=1}^m a_i z_i^k p_s(z_i)=\sum_{j=0}^{m-1}b_j\sum_{i=1}^m
	a_i z_i^{k+j} = \sum_{j=0}^{m-1}b_j w_{k+j},\label{limazp}
\end{equation}
where $w_k\isdef\sum_{i=1}^m a_i z_i^k$.
Since, for all $j=0,\ldots,m-1,$ $\lim_{k\to\infty}w_{k+j}=0$, it follows that
\begin{equation*}
	\lim_{k\to\infty}\sum_{i=1}^m a_i z_i^k p_s(z_i)=0.
\end{equation*}
Next, note that, since $p_s(z_s)=1$ and, for all $t \in \{1, \ldots, m\}\backslash\{s\},$  $p_s(z_t)=0,$ it follows that
\begin{equation}\label{eq:lemmApp2}
	{\sum_{i=1}^ma_i z_i^k p_s(z_i)} = {a_sz_s^k}.
\end{equation}
It thus follows from \eqref{limazp} that 
\begin{equation}
	a_s\lim_{k\to\infty}z_s^k=0.\label{aszsk}
\end{equation}

Now, suppose that $a_s\ne0.$  Then, \eqref{aszsk} implies that $\lim_{k\to\infty}|z_s|^k=0.$
Hence, $|z_s|<1,$ which contradicts the assumption that $|z_s|\ge1.$
Therefore, $a_s = 0,$ and thus $a_1=\cdots=a_m=0.$
\hfill $\square$

 The following is a corollary of Lemma \ref{appLem1}.

\begin{cor}\label{corApp1}
Let $z_1,\ldots,z_m\in\BBC$ be distinct, assume that, for all $i=1,\ldots,m,$ $|z_i|\ge1,$ and let $a_1,\ldots,a_m\in\BBC$, at least one of which is nonzero. Then,
\begin{equation}
	\limsup_{k\to\infty}\abs{\sum_{i=1}^m a_i z_i^k}>0.
\end{equation}
\end{cor}

The following result is used in the proof of Lemma \ref{propGyvalphaNS}.

\indent\begin{prop} \label{propA3}
Let $\lambda_1,\ldots,\lambda_n\in\BBC$ be distinct, and assume that $\max_{i=1,\ldots,n}|\lambda_i|>1.$
Furthermore, let $p_1,\ldots,p_n$ be nonzero polynomials with complex coefficients,
and, for all $k\ge1,$ define
\begin{equation}
    y_k\isdef\sum_{i=1}^n p_i(k)\lambda_i^k.
\end{equation}
Then,
\begin{equation} \label{eq:propA3Limsup}
    \limsup_{k\to\infty}\abs{y_k}=\infty.
\end{equation}
\end{prop}

{\bf Proof: }
Label $\lambda_1,\ldots,\lambda_n$ such that $|\lambda_1|\ge|\lambda_2|\ge\cdots \ge |\lambda_n|>0,$ and define 
$\rho\isdef\max_{i\in\{1,\ldots,n\}}\abs{\lambda_i}>1$ and $s\in\{1,\ldots,n-1\}$ such that
	$\rho=\abs{\lambda_1}=\cdots=\abs{\lambda_s}>	\abs{\lambda_{s+1}}.$
Furthermore, let $y_k=y_{k,1}+y_{k,2},$ where
\begin{equation} %
	y_{k,1}\isdef\sum_{i=1}^s p_i(k)\lambda_i^k, \quad
	y_{k,2}\isdef\sum_{i=s+1}^n p_i(k)\lambda_i^k,\\
\end{equation}
where, in the case $s = n,$ $y_{k,2}\isdef 0.$
Note that, for all $k\ge0,$
\begin{align*} 
    \abs{\frac{y_{k,2}}{\rho^k}}&=\abs{\sum_{i=s+1}^n p_i(k)
    \left(\frac{\lambda_i}{\rho}\right)^k} \\
    &\le \left(\sum_{i=s+1}^n\abs{p_i(k)}\right)
    \abs{\frac{\lambda_{s+1}}{\rho}}^k,
\end{align*}
which implies that
\begin{equation}\label{eq:propEq2}
    \lim_{k\to\infty}\frac{y_{k,2}}{\rho^k}=0.
\end{equation}
Next, let $\lambda_1,\ldots,\lambda_s$ be ordered such that
    $d \isdef \deg p_1=\deg p_2=\cdots=\deg p_m>\deg p_{m+1}\ge \cdots\ge\deg p_s,$
where $m\in\{1,\ldots,s\}.$
Thus, for all $i\in\{1,\ldots,m\},$ there exists a nonzero complex number $a_i$ and a polynomial $q_i$ such that $\deg q_i<d$ and, for all $k\ge0,$ $p_i(k)=a_i k^d + q_i(k).$ 
Hence,
\begin{equation}\label{eq:propEq3}
    \frac{y_{k,1}}{\rho^k k^d} = \sum_{i=1}^m a_i z_i^k+
    \sum_{i=1}^m \frac{q_i(k)}{k^d}z_i^k+
    \sum_{i=m+1}^s  \frac{p_i(k)}{k^d}z_i^k,
\end{equation}
where, for all $i \in \{1, \ldots, s\},$  $z_i\isdef\lambda_i/\rho.$
It thus follows from \eqref{eq:propEq3} that
\begin{equation}\label{eq:propEq4}
    \lim_{k\to\infty}\left(	
    \frac{y_{k,1}}{\rho^k k^d}-\sum_{i=1}^m a_i z_i^k\right)=0.
\end{equation}
Hence, \eqref{eq:propEq2} and \eqref{eq:propEq4} imply that
\begin{equation}\label{eq:propEq5}
	\lim_{k\to\infty}\left(\frac{y_k}{\rho^k k^d}-\sum_{i=1}^m a_i z_i^k\right)=0.
\end{equation}

Next, since $\lambda_1,\ldots,\lambda_m$ are distinct, it follows that $z_1,\ldots,z_m$ are distinct.
Corollary \ref{corApp1} thus implies that there exist $\varepsilon>0$ and an increasing sequence $(k_j)_{j\ge0}$ of positive integers such that, for all $j\ge0,$
\begin{equation} \label{eq:appSeqSum1}
    \abs{\sum_{i=1}^m a_i z_i^{k_j}}>2\varepsilon,
\end{equation}
and, from \eqref{eq:propEq5}, that there exists a nonnegative integer $j_0$ such that, for all $j\ge j_0,$
\begin{equation}\label{eq:appSeqSum2}
	\abs{\frac{y_{k_j}}{\rho^{k_j}k_j^d}-\sum_{i=1}^m a_i z_i^{k_j}}<\varepsilon.
\end{equation}
Hence, \eqref{eq:appSeqSum1} and \eqref{eq:appSeqSum2} imply that, for all $j\ge j_0,$
\begin{equation*}
	\abs{\frac{y_{k_j}}{\rho^{k_j}k_j^d}}\ge
	\abs{\sum_{i=1}^m a_i z_i^{k_j}}-
	\abs{\frac{y_{k_j}}{\rho^{k_j}k_j^d}-\sum_{i=1}^m a_i z_i^{k_j}}>
	\varepsilon,
\end{equation*}
and thus $\abs{y_{k_j}}> \varepsilon \rho^{k_j}k_j^d$.
Since $\rho>1$, it follows that
\begin{equation}
\lim_{j\to\infty}\abs{y_{k_j}}=\infty,\nn 
\end{equation}
which implies \eqref{eq:propA3Limsup}.  
\hfill $\square$

\begin{lem}\label{lemmalimits}

Let $M$ be an $n\times n$ real matrix, assume that 1 is not an eigenvalue of $M,$ and let $x$ and $y$ be $n\times 1$ real vectors.
Then the following statements hold:
\begin{enumerate}
    \item If $\lim_{k\to\infty}M^k$ exists, then the limit is zero.
    \item If $\lim_{k\to\infty}M^kx$ exists, then the limit is zero.
    \item If $\lim_{k\to\infty}y^\rmT M^kx$ exists, then the limit is zero.
\end{enumerate}
\end{lem}

{\bf Proof: }
To prove {\it iii}), write  $\chi_M(z) = z^n+a_1z^{n-1}+\cdots+a_n$, and, for all $k\ge0,$ define
$t_k\isdef y^\rmT M^k x$.  Then the Cayley-Hamilton implies that, for all $k\ge0,$
\begin{equation}
t_{n+k}+a_1t_{n+k-1}+\cdots+a_nt_k=0.\label{tkeqn}
\end{equation}
Since $\lim_{k\to\infty}t_k$ exists, letting $k\to\infty$ in \eqref{tkeqn} yields
\begin{align}
0&=(1+a_1+\cdots+a_n)\lim_{k\to\infty}t_k\nn\\
&=\chi_M(1)\lim_{k\to\infty}t_k.\label{limprod}
\end{align}
Since $\chi_M(1) \ne0,$ \eqref{limprod} implies that $\lim_{k\to\infty}t_k = 0.$
Finally, {\it i}) and {\it ii}) follow by taking $x$ and $y$ to be columns  of the $n\times n$ identity matrix.
\hfill $\square$



\end{appendices}

\section*{CRediT authorship contribution statement}

\textbf{Juan A. Paredes:} Conceptualization, Formal analysis, Writing
– original draft, Writing – review \& editing.
\textbf{Syed Aseem Ul Islam:} Conceptualization, Writing
– original draft.
\textbf{Omran Kouba:} Formal analysis, Writing – review \& editing.
\textbf{Dennis S. Bernstein:} Writing
– original draft, Writing – review \& editing, Supervision, Funding acquisition.

\section*{Declaration of competing interest}

The authors declare that they have no known competing financial interests or personal relationships that could have appeared to influence the work reported in this paper.

\section*{Acknowledgements}

The first author was supported by NSF grant CMMI 1634709.


\bibliographystyle{elsarticle-num-names}
\bibliography{main_DTL_arxiv}

\end{document}